\batchmode
\documentclass[english]{article}
\usepackage{graphicx}
\usepackage{amssymb}
\usepackage{lmodern}
\usepackage[T1]{fontenc}
\usepackage{geometry}
\usepackage[authoryear]{natbib}

\begin{document}

\title{Interval Estimation for the `Net Promoter Score'}

\author{Brendan Rocks}

\maketitle

Satmetrix Systems Inc., 3 Twin Dolphin Dr, Redwood
City, CA 94065 \\ e-mail: \texttt{rocks.brendan@gmail.com}
\\
\begin{abstract}
The Net Promoter Score (NPS) is a novel summary statistic used by
thousands of companies as a key performance indicator of customer
loyalty. While adoption of the statistic has grown rapidly over the
last decade, there has been little published on its statistical properties.
Common interval estimation techniques are adapted for use with the
NPS, and performance assessed on the largest available database of
companies' Net Promoter Scores. Variations on the Adjusted Wald, and
an iterative Score test are found to have superior performance.

\end{abstract}

\section{The Net Promoter Score}

\subsection{Usage and Calculation}

The Net Promoter Score (NPS) is a summary statistic proposed by Reichheld
\citeyearpar{reichheld2003one,reichheld2006ultimate}, commonly used
in commercial survey research to estimate the propensity of a business'
customers to exhibit desirable behaviors, such as recommending friends,
or spending a greater share of their income (Owen \& Brooks, \citeyear{owen2008answering};
\citealp{reichheld2011ultimate}). General practice is to ask the
question \emph{``How likely is it that you would recommend Company
X to a friend or colleague?}'', with responses captured on a 0 to
10 Likert scale. The NPS statistic is then calculated as follows;
respondents who rate 0 to 6 are classified as \emph{Detractors}, 7
or 8 as \emph{Passives}, and 9 or 10 as \emph{Promoters}. The NPS
is calculated as the percentage of \emph{Promoters}, less the percentage
of \emph{Detractors}, producing a score between -1 and 1. \footnote{Net Promoter Scores are often multiplied by 100 (and occasionally
accompanied by a percentage sign) for presentational purposes, although
this is omitted in this paper.}

We'll consider the number of respondents in each category a vector
of length three, $x=[x_{det},\:x_{pas,\:}x_{pro}]$, with their relative
proportions the corresponding probability vector $p=[p_{det},\:p_{pas,\:}p_{pro}]$,
the score itself being $NPS=p_{pro}-p_{det}$. The score may also
be reached by recoding \emph{Promoter, Passive} and \emph{Detractor}
responses as 1, 0, and -1, respectively, and taking the arithmetic
mean.

This paper focuses on estimating intervals for the \emph{$NPS$} statistic
itself, as opposed to other measures which might describe the trinomial
distribution used to derive it. This is an important distinction;
a single $NPS$ can come from many (potentially rather different)
distributions.

\subsection{Critiques}

A variety of metrics thought to predict customer behaviors exist within
marketing, and Riechheld's (\citeyear{reichheld2003one}) claim that
the $NPS$ is superior has been challenged by several authors. In
particular, on the grounds that $NPS$ and alternative metrics have
similar relationships to business-outcomes (\citealt{van2013satisfaction,pingitore2007,keiningham2007a});
that an 11-point Likert scale may not be the optimal measurement instrument
(\citealt{schneider2008}); and that multiple measures combined and
weighted via a regression model provide better predictions (\citealt{keiningham2007b}).
Compared to taking the mean on the original scale, the novel calculation
has been argued to both lose information (\citealt{eskildsen2011}),
and improve performance in predicting customer retention (\citealt{de2015predictiv}).
Despite these critiques, the Net Promoter Score is used to estimate
customer sentiment by thousands of companies (\citealt{owen2008answering,reichheld2011ultimate}).
This paper investigates its statistical properties.

\subsection{Properties}

Many possible trinomial probability mass distributions (TPMDs) can
result in an $NPS$ of 0, half that number for an $NPS$ of $\frac{1}{2}$,
and only 1 for an $NPS$ of 1 (or -1). For any $n$, there are $2n+1$
possible Net Promoter scores, the distribution having a peak 1 score
wide at 0, with $n$ `steps' of two scores width either side for even
$n$, and a peak of 3 scores wide, with $n-1$ steps for odd numbered
$n.$

Unlike the {[}0,1{]} uniform distribution of possible values of a
binomial proportion, possible values of the $NPS$ from a simplex
lattice follow a triangular distribution ($a=1,b=-1,c=0$) as $n$
approaches infinity.

\begin{figure}

\includegraphics[width=6.5in]{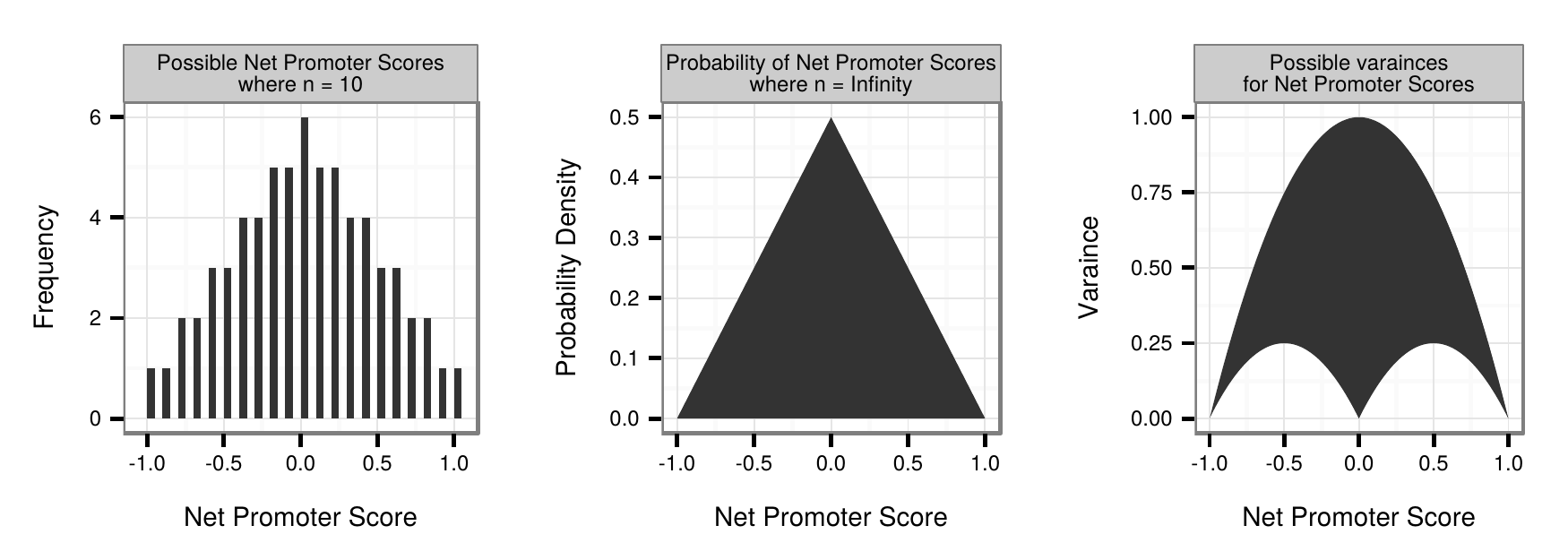}

\caption{Illustrations of Net Promoter Scores drawn from simplex lattices.
The left panel illustrates the discrete distribution of possible Net
Promoter Scores for $n=10$. The center panel shows the smooth triangular
distribution when $n$ approaches infinity - the binomial equivalent
would be uniform. The rightmost panel shows the range of possible
Net Promoter Scores and variances for infinite $n$, the possible
distributions being uniformly distributed within the \emph{inverted
shield} shape.\label{fig:properties}}
\end{figure}

This is an important distinction with regard to assessing interval
methods; performance averaged uniformly across possible TPMDs is not
performance averaged uniformly over Net Promoter Scores (Figure \ref{fig:properties}).

Testing a method for $NPS$ with equal weight across TPMDs, means
that (for example) performance at $NPS=0$ will have twice the weight
of performance at $NPS=\frac{1}{2}$, as for arbitrary trinomial distributions,
an $NPS$ of $0$ is twice as likely to occur.

\subsection{Variance of the NPS}

Methods for the variance of the difference of two proportions can
be applied to the $NPS$ (e.g. \citealp{gold1963tests}; \citealp{goodman1965simultaneous})\footnote{More recently, an alternative, but equivalent derivation, for the
variance of the $NPS$ was published online by \citet{whuber}.}
\[
\sigma_{NPS}=p_{pro}+p_{det}-(p_{pro}-p_{det})^{2},
\]
with the variance ranging from 0 (all respondents in the same category,
for example \emph{Passives}), to a maximum of 1 (data equally split
between \emph{Promoters} and \emph{Detractors}). It's worth noting
that these two extreme examples would both produce an $NPS$ of 0;
unlike a binomial proportion, we cannot derive an $NPS$ from its
variance.

\section{Interval estimation}

\subsection{Wald Intervals, and Variations}

\subsubsection{The Wald Interval}

A commonly taught and used method for sample proportions is the Wald
confidence interval (first proposed by Laplace, \citeyear{de1820theorie}),
$p\pm z_{\alpha/2}\sqrt{p(1-p)/n}$, where $z_{\alpha/2}$ denotes
the $1-(\alpha/2)$ quantile of a standard normal distribution. It
is straightforward to use the variance calculation (1) to produce
a Wald interval for the $NPS$:

\begin{equation}
NPS\pm z_{\alpha/2}\sqrt{\frac{\sigma_{NPS}}{n}}.\label{eq:wald}
\end{equation}

\subsubsection{The Goodman method}

\citet{goodman1964simultaneous}, proposed a method for estimating
net differences between multinomial parameters. It functions in a
similar form to the Wald interval, with the sample $NPS$ forming
the central point of the interval
\[
\pm\sqrt{\chi\frac{\sigma_{NPS}}{n}}
\]
 where $\chi$ is the upper $(a/K)\ensuremath{\times}100th$ percentile
of the $\chi^{2}$ distribution with one degree of freedom.

\subsubsection{The Adjusted Wald}

The `Adjusted Wald' test proposed by Agresti \& Coull (\citeyear{agresti1998approximate})
in its original binomial form is to perform the Wald test, after the
adjustment of adding $\frac{z_{\alpha/2}^{2}}{2}$ to the number of
successes, and $z_{\alpha/2}^{2}$ to the number of trails. Similarly,
Agresti \& Min \citeyearpar{agresti2005simple} proposed an Adjusted
Wald for matched pairs in $2\times2$ contingency table designs.

We can adapt this to the $NPS$ by adding $\frac{z_{\alpha/2}^{2}}{3}$
to the number of respondents in each category, so that $\hat{x}=x+\frac{z_{\alpha/2}^{2}}{3}$,
and $\hat{n}=n+z_{\alpha/2}^{2}$, making our adjusted estimate of
the TPMD $\hat{p}=\frac{\hat{x}}{\hat{n}}$, the new central estimate
$\widehat{NPS}=\hat{{p}}_{pro}-\hat{{p}}_{det}$, and new variance
$\hat{\sigma}_{NPS}=\hat{{p}}{}_{pro}+\hat{{p}}_{det}-(\hat{{p}}_{pro}-\hat{{p}}_{det})^{2}$.
We then use these adjusted parameters to create intervals using the
Wald method in \ref{eq:wald} above:

\[
\widehat{NPS}\pm z_{\alpha/2}\sqrt{\frac{\hat{\sigma}_{NPS}}{\hat{n}}.}
\]
These adjustments shrink the estimated TPMD towards the uniform, the
additions to $\hat{x}$ bringing $\widehat{NPS}$ closer to 0, and
the estimated variance closer to $\frac{2}{3}$.

The weights added to $x$ need not necessarily sum to $z_{\alpha/2}^{2}$,
or be equally distributed across the trinomial categories. Agresti
\& Coull \citeyearpar{agresti1998approximate} proposed a total weight
of 4 (as opposed to $z_{\alpha/2}^{2}$). Agresti \& Min's (2005)
specification for matched-pairs advocates adding the same weight to
each of the four categories in a $2\times2$ table, which when respecified
for a TPMD, can be considered adding twice the weight to $p_{pas}$
than is added to $p_{pro}$ and $p_{det}$. This does not affect the
central estimate of the interval, but has the effect of reducing the
estimated variance and interval width.

Bonett \& Price \citeyearpar{bonett2012adjusted} suggested another
novel adjustment for the Wald, again in the context of matched pairs
and $2\times2$ tables, which is to add the weight to just the cells
subject to the statistic's calculation - in our case, the weight split
equally between the trinomial extremes of $p_{pro}$ and $p_{det}$.

\paragraph{Notation for Adjusted Wald Interval Estimates}

This paper uses the notation $AW(w,shape)$ to denote an Adjusted
Wald interval, where $w$ is the total weight added to $\hat{x}$,
and $shape$ can be extreme (E), triangular (T), or uniform (U); denoting
$p_{pas}$ having no weight, twice the weight, or the same weight
as the other categories, respectively. This results in the prior having
a variance of $1$ (for E) $\frac{2}{3}$ (for U) or $\frac{1}{2}$
(for T). For example, an Adjusted Wald interval with one response
added to each trinomial category would be denoted $AW(3,U)$. This
paper assesses Adjusted Wald 95\% intervals where $w$ is equal to
$2$, $3$, and $z_{\alpha/2}^{2}$ ($\approx3.84$), for all three
shape types.

\subsection{Score Tests}

\subsubsection{The Score Test}

The score test, originally proposed by \citet{wilson1927probable}
has the binomial formula

\begin{equation}
\left(p+\frac{z_{\alpha/2}^{2}}{2n}\pm z_{\alpha/2}\sqrt{\left[p(1-p)+z_{\alpha/2}^{2}/4n\right]/n}\right)/(1+z_{\alpha/2}^{2}/n).\label{eq:wilson}
\end{equation}
As presented in Agresti \& Coull's illuminating \citeyear{agresti1998approximate}
paper, the central point of the interval can be alternatively specified
as a weighted average, $p(w_{1})+\frac{1}{2}(w_{2})$, the two weights
being $w_{1}=\frac{n}{n+z^{2}}$ and $w_{2}=\frac{z^{2}}{n+z^{2}}$
respectively. This weighted average shrinks $\hat{p}$ towards $\frac{1}{2}$,
with this effect diminishing as $n$ increases. Standard errors either
side of this midpoint are $z_{\alpha/2}\sqrt{\frac{p(1-p)w_{1}+\frac{1}{4}w_{2}}{n+z^{2}}}$,
providing a weighted average between the sample variance, and the
maximum possible variance of $\frac{1}{4}$.

Using the the weighted average principle, we can adapt this to the
$NPS$, with the two weights shrinking the central estimate towards
0 as opposed to $\frac{1}{2}$, as follows:

\[
\widehat{NPS}=\left(NPS+1\right)w_{1}+w_{2}-1,
\]
The formula for the intervals,
\begin{equation}
\widehat{NPS}\pm z_{\alpha/2}\sqrt{\frac{\sigma_{NPS}w_{1}+w_{2}}{n+z^{2}}}\label{eq:score_interval}
\end{equation}
 is similar in form to the original Wilson score test, but with the
weighted average drawing the variance towards the $NPS$ maximum of
1. This prior variance can be altered by the addition of a multiplier
to $w_{2}$; in this paper prior variances of $\frac{2}{3}$ and $\frac{1}{2}$
are tested, to provide equivalence with the prior variances of the
uniform and triangular Adjusted Wald tests.

\subsubsection{The Iterative Score Method}

Inverting the score test was first proposed for paired sample designs,
and applications to $2\times2$ tables by \citet{tango1998equivalence}.
This test can be reinterpreted to cover the $NPS$ of a trinomial
distribution. Modifying Agresti \& Min's \citeyearpar{agresti2005simple}
presentation, its interval would be the set of values $\Delta$, satisfying

\[
\frac{|(NPS)-\Delta|}{\sqrt{\frac{(\tilde{p}_{pro}(\Delta)+\tilde{p}_{det}(\Delta))-\Delta^{2}}{n}}}<z_{\alpha/2}
\]
where $\tilde{p}_{i}(\Delta)$ is the MLE of $p_{i}$, under the constraint
$\Delta=NPS$. This can be solved iteratively; for this paper, the
implementation was adapted from code for the \citet{tango1998equivalence}
method by \citet{agresti2003tangocode}.

\subsubsection{The May-Johnson Score Method}

May \& Johnson \citeyearpar{may1997confidence} proposed a closed
form version of Tango's method, again originally intended for $2\times2$
tables from matched pairs designs. It can be adapted to trinomial
data and the \emph{$NPS$} as follows

\[
NPS\left(\frac{n}{\hat{n}}\right)\pm z_{\alpha/2}\sqrt{\frac{\hat{n}(p_{pro}+p_{det})-n(p_{pro}-p_{det})^{2}}{\hat{n}}}
\]
where $\hat{n}=n+z_{\alpha/2}^{2}$.

\subsection{Similarity Between Methods}

The Score Method, May-Joshnson Score Method, and Adjusted Wald tests
with a weight of $z_{\alpha/2}^{2}$, all produce identical central
estimates for the interval. Both the Goodman and Wald methods take
the sample $NPS$ as the central estimate.

\begin{figure}

\includegraphics[width=6.5in]{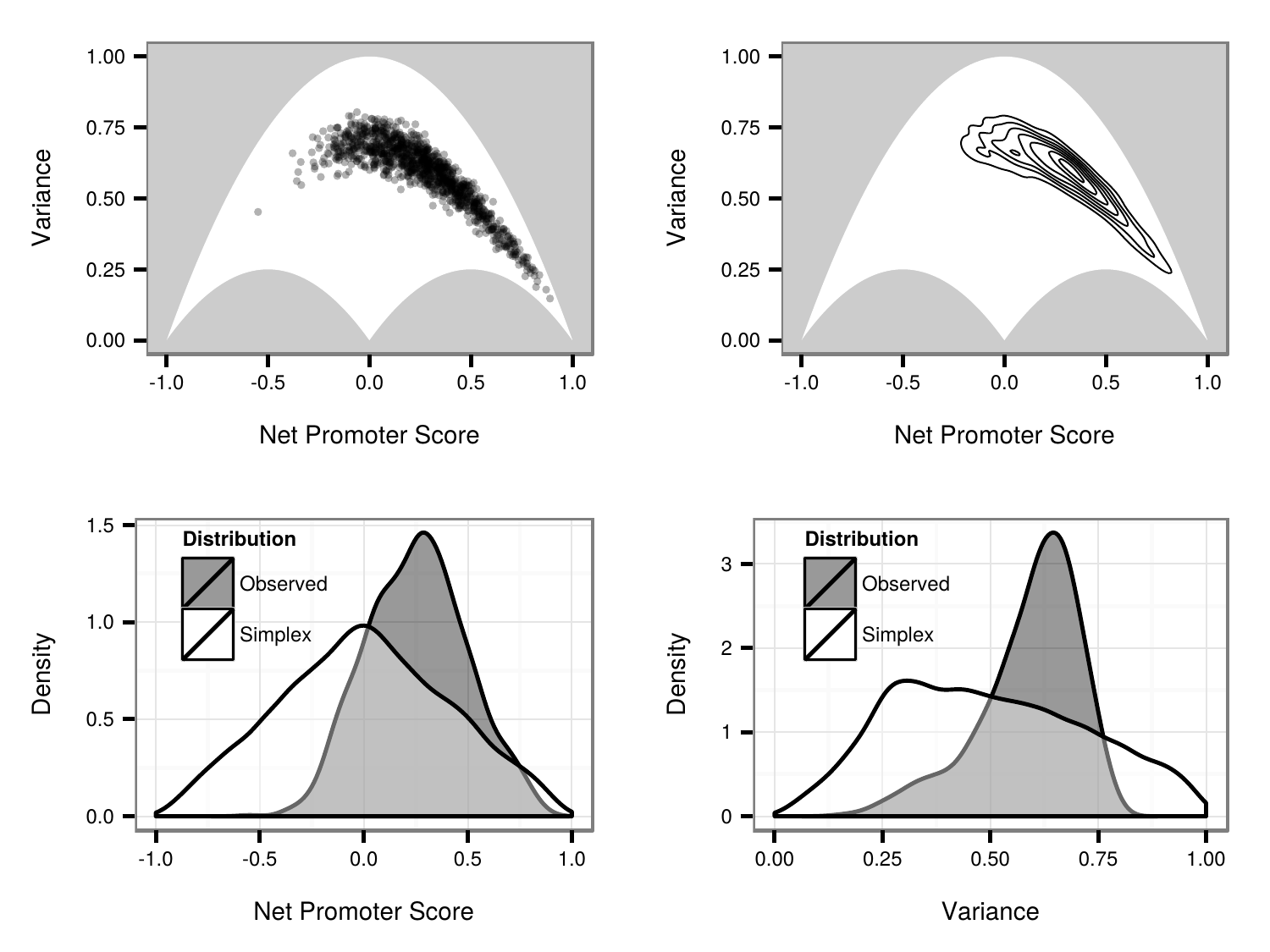}

\caption{The $NPS$ and variance of the 1,098 observed TPMDs from the Satmetrix
data set, illustrated with a scatter plot (upper left panel), and
contour plot (upper right panel) of the two-dimensional kernel density
estimate. The area outside the range of possible distributions is
shaded gray. Marginal density estimates from the same model for the
$NPS$ (lower left panel), and variance (lower right panel) are also
shown, compared to those from the samples from the simplex lattice.
Compared to the range of possible TPMDs, those observed have higher
mean Net Promoter Scores and variances, and are much more tightly
grouped.\label{fig:obs-plots}}
\end{figure}

\section{Assessment of Coverage Probabilities}

\subsection{Methods}

The methods of Agresti \& Coull \citeyearpar{agresti1998approximate}
inform the simulation based approach for coverage probability assessment.
The specified confidence level of a procedure is compared to the long
run average of times that a procedure's interval contains the `true'
population parameter, when supplied data from a random sample of the
population. For the this analysis, the nominal confidence level chosen
is 95\%. This means that the results indicate \emph{average}, as opposed
to \emph{worst possible} performance; procedures where coverage probabilities
are greater than the nominal confidence level will be seen as overly
conservative, those with lower than nominal coverage probabilities
will be seen as overly liberal.

\subsubsection{Arbitrary Trinomial Distributions}

Trinomial probability mass distributions were generated by randomly
sampling $J=10,000$ points from a (3, 400) simplex lattice. Performance
at each TPMD was assessed at 20 $n$ counts\footnote{$n=$ 5 to 100 in intervals of 5. Additionally, performance at $n=$
120 to 300 (in intervals of 20) was assessed for descriptive purposes,
but not not used in the final selection criteria.}; \ensuremath{2\times 10^{5}} trinomial distributions in total. Performance
at each trinomial distribution was assessed with 10,000 simulations.
This is a sample of trinomial distributions from those which are \emph{arbitrarily
possible}.

\subsubsection{Observed Trinomial Distributions}

While sampling from a simplex lattice gives a good indication of performance
over\emph{ possible} distributions, in psychometric practice, some
distributions are more likely than others. The Satmetrix US Consumer
Net Promoter Study \citep{satm} is the largest available database
of companies' Net Promoter Scores. Aggregating at the interaction
of year-of-response and company, 347,788 Likelihood to Recommend ratings
for 236 companies over 14 years yielded 1,098 trinomial Net Promoter
distributions (with at least 250 responses). The data illustrate that
samples from a simplex lattice are not an ideal model of human response
behaviors (Figure \ref{fig:obs-plots}). The observed TPMDs have much
more narrowly distributed Net Promoter Scores (mean = .26, standard
deviation = .24) and variances (mean = .59, standard deviation = .12)
than the simplex lattice samples, and occupy a relatively small small
area of the possible parameter space.

\paragraph{Performance more likely to be observed in practice}

To create statistics which reflect performance across values sampled
from the simplex lattice, performance is averaged across the $J$
TPMDs sampled from it. For statistics which might better reflect performance
\emph{in practice}, we can make this a weighted average, the weights
reflecting how frequently such a TPMD has been observed. To create
these weights, a two-dimensional kernel density estimate was fit to
the $NPS$ and variance of the trinomial distributions observed in
the Satmetrix data-set \footnote{Bivariate kernel density estimate fit using pilot bandwidth selection
(Chac{\'o}n \& Duong, \citeyear{chacon2010multivariate}), resulting
in 151 evaluation points) via the \texttt{R }package \texttt{ks} \citep{ks}. }, the weights being the density estimate of a given distribution (rescaled
so that the sum of the weights across the $J$ samples is 1). This
paper presents performance both with and without these observational
weights applied.

\subsubsection{Desirable Performance Characteristics}

In addition to a test having an average coverage level close to 95\%,
the following characteristics are desirable:
\begin{itemize}
\item Good performance across values for $n$, especially $\leqslant100$
\item Low \emph{variation} in performance across trinomial distributions.
For example, a test may have an \emph{average} coverage probability
of 95\%, by returning extremely conservative results for certain distributions,
and extremely liberal results for others
\item Good performance for both the observed and simplex distributions
\end{itemize}
A convenient summary of these properties is the mean absolute error
(MAE) of the test, defined at a particular $n$ value, as
\[
MAE=\frac{\sum(C_{j}-0.95)w_{j}}{\sum w_{j}}
\]
where $C_{j}$ is the coverage probability of the test for the $j^{th}$
TPMD sampled from the simplex lattice, and $w_{j}$ is the weight
for that distribution. The tests' MAE for $n\leqslant100$ will be
used as our ultimate criteria for recommendation. This paper considers
performance both with, and without ($w_{1,\ldots,J}=\frac{1}{J}$)
observational weights applied to the MAE.

\begin{table}

\includegraphics[width=6.5in]{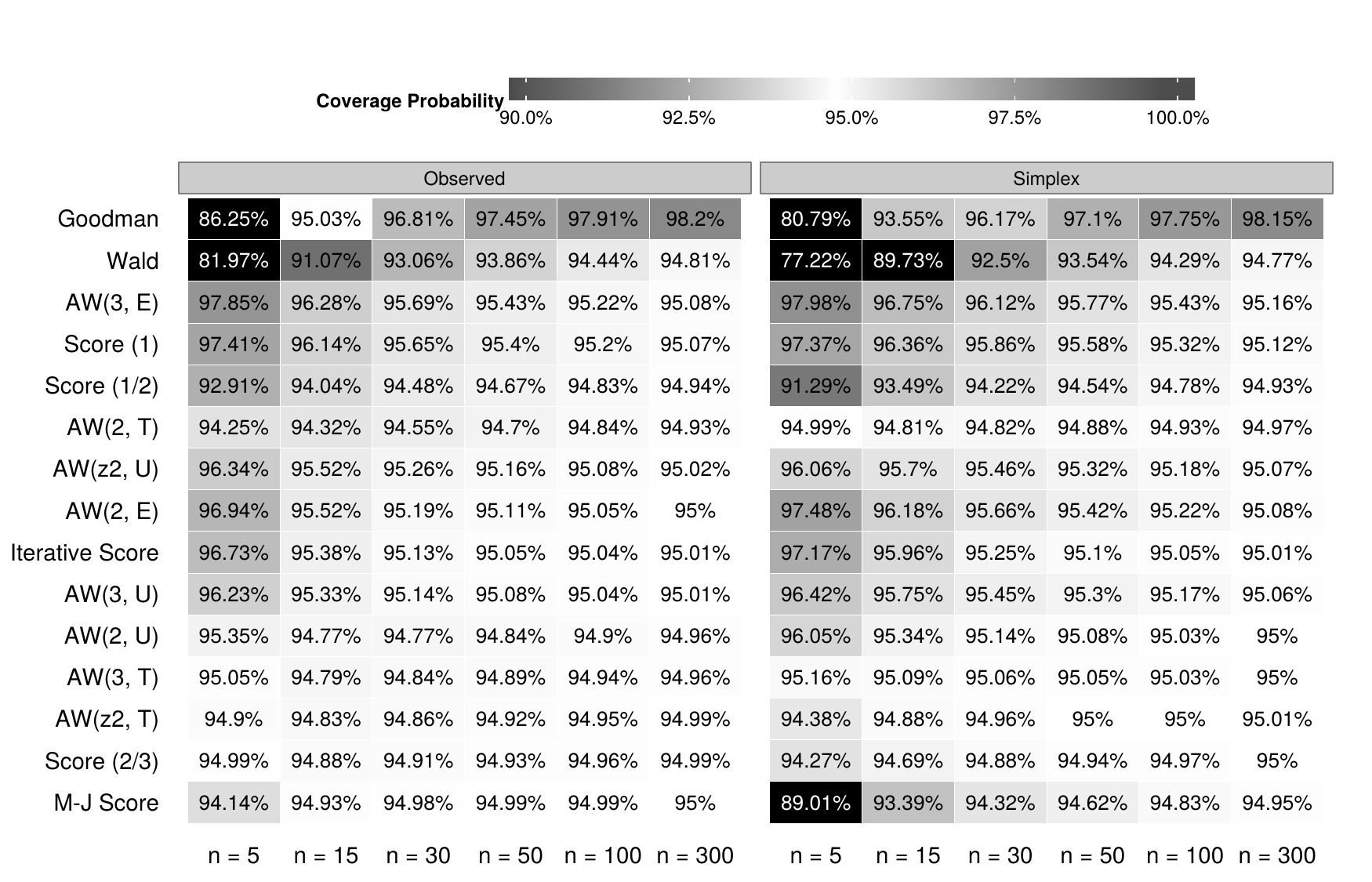}

\caption{A coverage probability heat map for the different interval estimation
methods varying with sample $n$ counts, for the observed (left panel)
and simplex (right panel) distributions. Tests are ordered by average
coverage probability for the observed distribution, where $n\leqslant100$.
Coverage probabilities below $90\%$ are filled solid black with white
text.\label{tab:ff-cov-tab}}
\end{table}

\subsection{Results}

\begin{figure}

\includegraphics[width=6.5in]{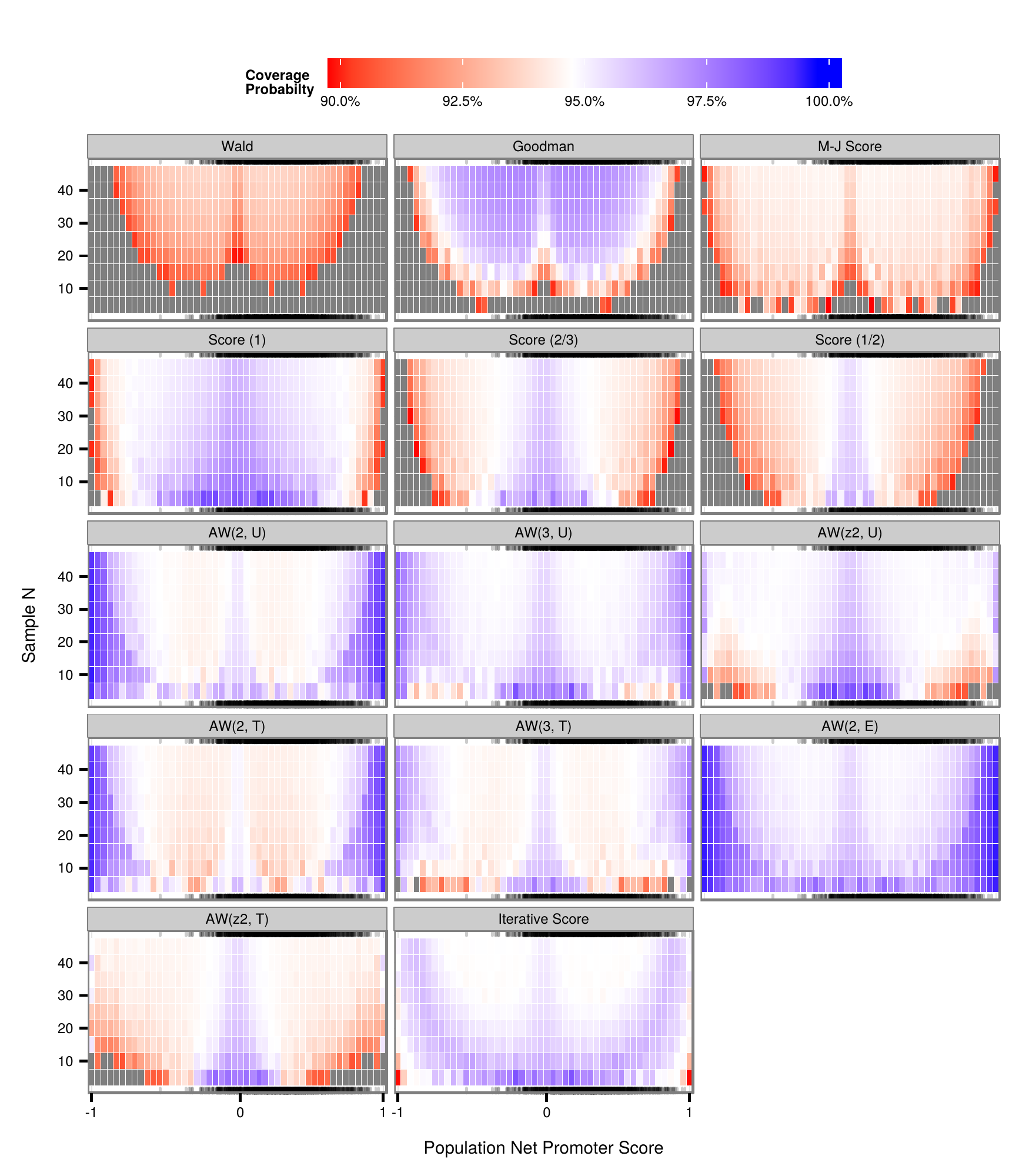}

\caption{A coverage probability heat map for the different interval estimation
methods, with varying Net Promoter Scores and $n$ counts. Rug plots
above and below each panel illustrate the observed distribution of
Net Promoter Scores. Coverage probabilities below $90\%$ are shaded
gray.\label{fig:ff-pool-plot}}
\end{figure}

\begin{figure}

\includegraphics[width=6.5in]{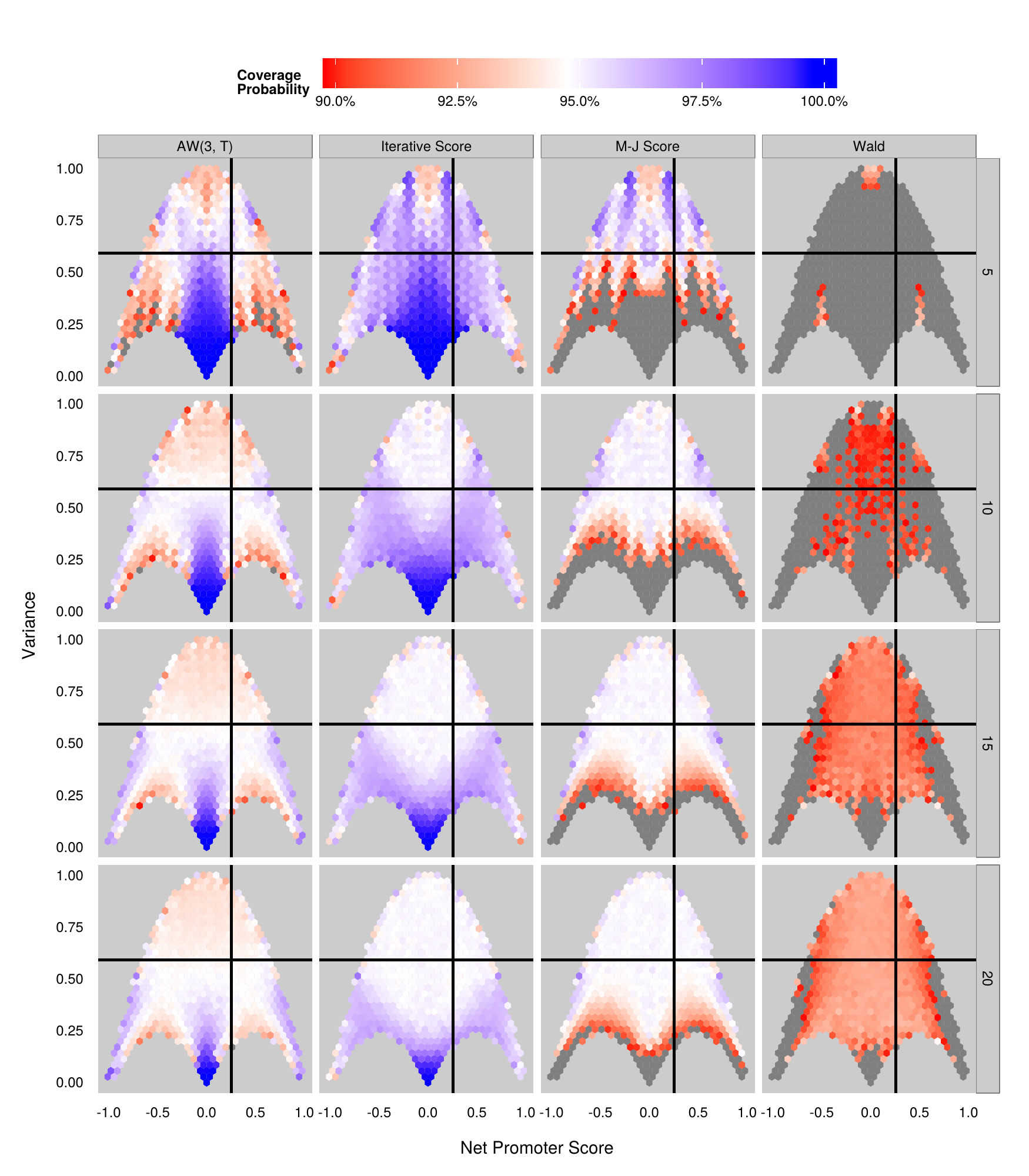}

\caption{Coverage probabilities for the $AW(3,T)$, Iterative Score, May-Johnson
Score, and Wald methods, across the TPMD parameter space, for varying
$n$. The mean $NPS$ and variance of the observed distribution is
indicated on each small multiple with cross-hairs. Areas with coverage
probabilities below 90\% are shaded dark gray. Note that for the $AW(3,T)$,
the area around the mean of the bivariate distribution is closest
to its region of optimal performance, compared to the Iterative Score,
where it is in a region of conservative bias.\label{fig:shield}}
\end{figure}
Table \ref{tab:ff-cov-tab} shows the average coverage probabilities
for the tests at 95\% intervals. The tests share the same general
characteristic of performance closer to the nominal level with increasing
$n$, the exception being the Goodman method. It produces intervals
which are too small with low $n$, and to wide with large $n$, the
test passing through the nominal coverage level at $n$ of around
20 for the simplex distribution.

The Wald test for $NPS$ fares better than its binomial equivalent
in Agresti \& Coull \citeyearpar{agresti1998approximate}, though
performance is still overly liberal. Coverage improves with increasing
$n$, though doesn't quite reach the nominal 95\% level by $n=300$.

\begin{table}

\includegraphics[width=6.5in]{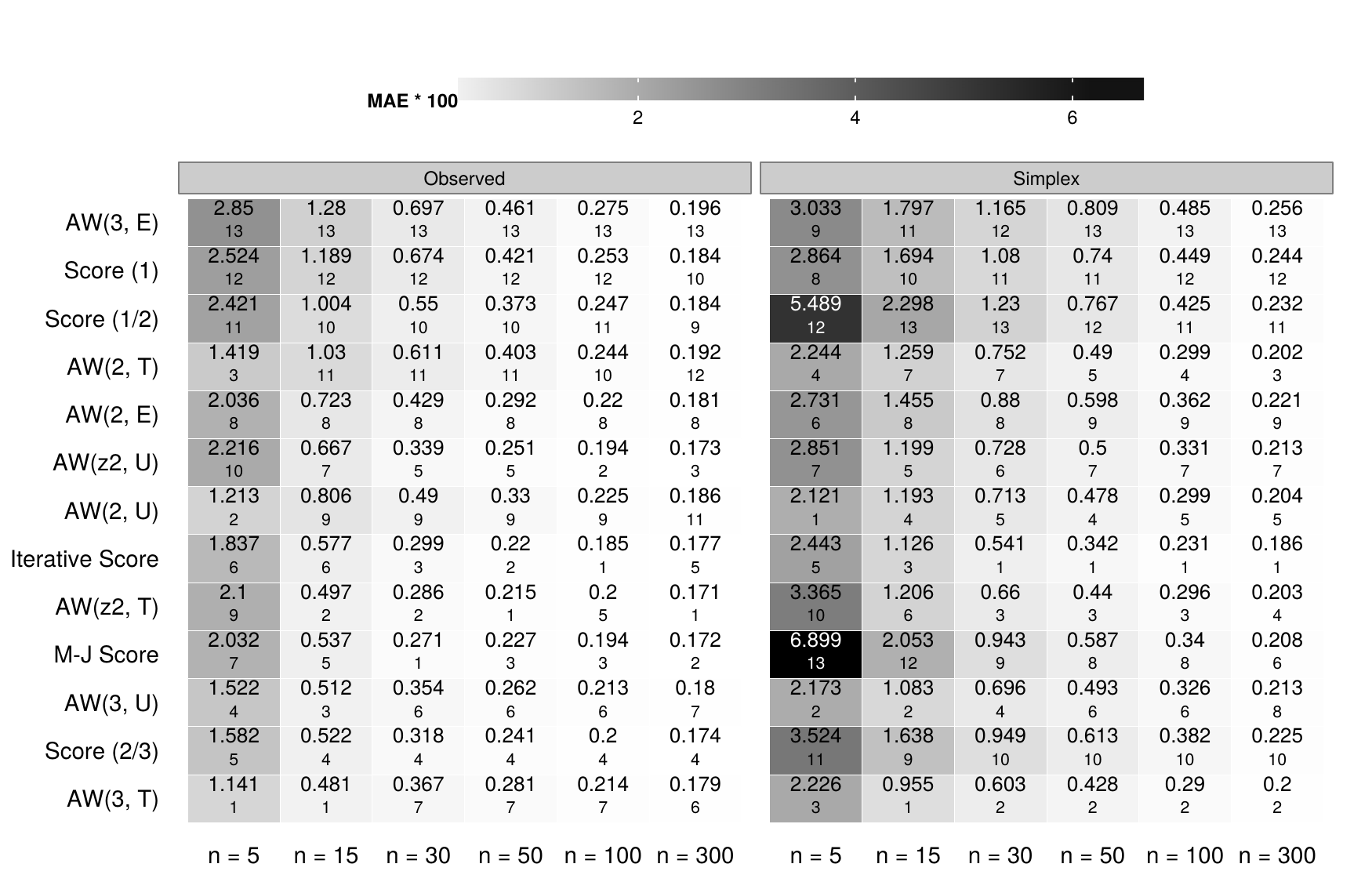}

\caption{Mean Absolute Error (multiplied by 100 for presentational purposes)
and rank, for interval estimation methods, for both the simplex and
observed distributions. Tests are ordered by descending total MAE
for the observed distribution, where $n\leqslant100$. \label{tab:rmse}}
\end{table}

The May-Johnson Score test has the best \emph{average} coverage probability
for the observed distribution. However, this comes at the expense
of \emph{variation} in performance (illustrated in Figure \ref{fig:ff-pool-plot}),
which is rather high, the coverage probability falling to below 90\%
at $n=5$ on the simplex distribution, and being overly liberal at
extreme and central $NPS$ values at low $n$.

\paragraph{Weighted Average Variations on the Score Test}

For the `weighted average' variations on the Wilson type score tests,
drawing the variance towards $\frac{1}{2}$ and $\frac{2}{3}$ produces
an on-average improvement in coverage probability over the original
specification of a prior variance of 1 (Table  \ref{tab:ff-cov-tab}).
Unfortunately, this comes at the expense of greater variance in performance
across Net Promoter Scores at low $n$ (Figure \ref{fig:ff-pool-plot}).

\paragraph{Variations on the Adjusted Wald}

All Adjusted Wald methods are superior to the Wald, with the $AW(3,T)$
the best of those tested, on both the observed and simplex distributions.
For the observed distribution, it has the lowest MAE for $n\leqslant100$
of any test, and the second lowest for the simplex (Table \ref{tab:rmse}),
very little coverage below the nominal level (Figure \ref{fig:ff-pool-plot}),
and superb average coverage probabilities (Table \ref{tab:ff-cov-tab}),
for both the observed and simplex distributions.

\paragraph{The Iterative Score}

The Iterative Score method also has excellent performance. Summing
MAE for all $n$, it has the lowest total for the simplex distribution,
and its performance is very rarely over-liberal, the test returning
coverage below 90\% the least frequently of any considered (less than
0.01\% of simplex samples), and having the highest minimum coverage
observed in the simulations (83\%). However, it has the disadvantage
of being overly conservative at low n. Like the Adjusted Wald tests,
it's more conservative at the extremes of $NPS$ for small $n$.

\paragraph{Performance with varying \emph{n}}

Our main performance statistic for recommending a test (MAE for $n\leqslant100$)
contains an intentional bias, in that it favors tests which have better
performance at low $n$ values, where MAE tends to be higher, making
a greater contribution to the aggregate. The $AW(3,T)$ benefits from
this the most, having better relative performance for observed and
simplex distributions at $n$ below around 15 and 20, respectively.

An alternative statistic might be to rank our tests' performance at
each value of $n$, and then select the method with the lowest average
rank. For the simplex distribution, using this criteria makes little
difference, with the Iterative Score, followed by the $AW(3,T)$ having
the lowest average rank. However, for the observed distribution, the
$AW(3,T)$ falls to eighth place, with the May-Johnson score coming
out the best, followed by the $AW(z_{\alpha/2}^{2},T)$.

\begin{figure}

\includegraphics[width=6.5in]{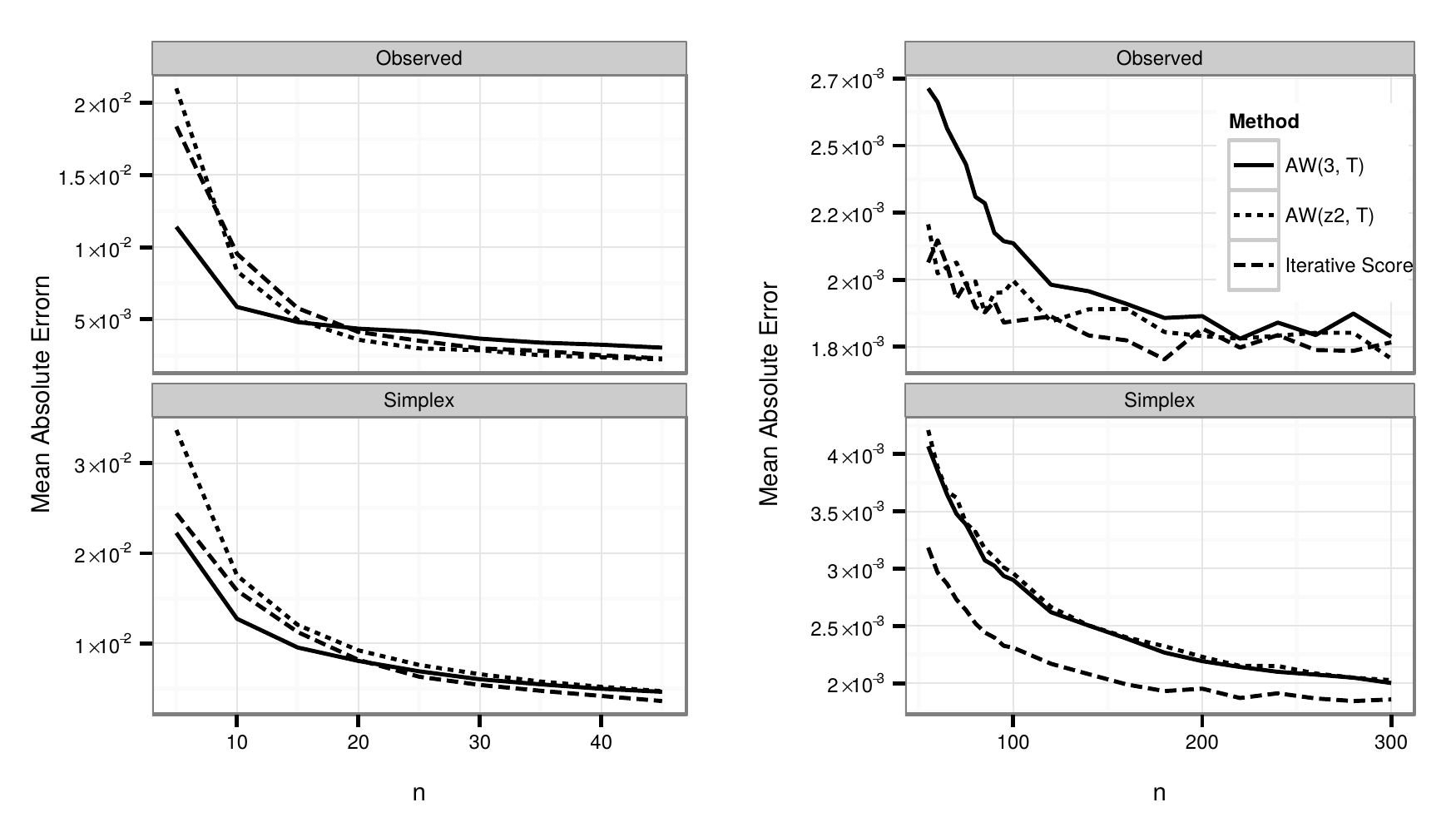}

\caption{Mean Absolute Error with varying $n$, for the Iterative score, AW(3,
T), and AW$(z_{\alpha/2}^{2}$, T) tests.\label{fig:ac-mae}}
\end{figure}

Of course, $n$ counts are known at the time of interval construction;
it is possible to select the best performing test at any of the intervals
for $n$ analyzed in this paper, or create a single `method' where
the underlying calculations change based on the total $n$. However,
doing so offers extremely modest performance improvements (reducing
the MAE by \ensuremath{4.09\times 10^{-4}} and \ensuremath{5.38\times 10^{-4}}
for the simplex and observed distributions respectively).

The difference between the MAEs of the tests decreases rapidly with
$n$, meaning that there is very little difference in MAE between
the ranks after around $n=30$. Figure \ref{fig:ac-mae} illustrates
this pattern, the $AW(3,T)$ having superior performance, when differences
in performance are of the highest magnitude.

\paragraph{Observed vs. Simplex distributions}

Our choice of two distributions provides us with two different sets
of results to judge our tests by. It's clear from the observational
data (Figure \ref{fig:obs-plots}) that performance in a relatively
small region of parameter space is of much greater importance under
the conditions observed. Figure \ref{fig:shield} illustrates the
overlap of observational data and coverage probability in parameter
space, explaining the large increase in performance in the $AW(3,T)$
and May-Johnson Score tests that are seen on the observed vs. simplex
distributions.

The observations have been selected to be representative of US Consumers
using the standard survey methodology, and thus perhaps the data generating
mechanism from which practitioners are most likely to encounter the
$NPS$. However, response behaviors are known to vary by both industry
and country (Owen \& Brooks, \citeyear{owen2008answering}), and interval
estimation methods may be applied to `net proportion' statistics outside
of traditional data collection for a Net Promoter Score. Performance
on both distributions are presented here, and it is left to the reader
to select the best test for their particular application.

\subsection{Additional Confidence Levels}

While the 95\% confidence interval is perhaps the most commonly used,
an important consideration for a test is performance at a range of
common confidence levels. The analysis above was replicated for a
subset of tests (the Iterative Score, the Score $\left(\frac{2}{3}\right)$,
the $AW(3,T)$, and the $AW(z^{2},T)$), for 99\%, 90\% and 80\% confidence
intervals\footnote{Tested at $n$ from 5 to 100 in intervals of 5.}.
The tests represent the best performing test of each type (closed-form
score, Adjusted Wald, and Iterative score) at 95\% confidence, with
the addition of the $AW(z^{2},T)$, which varies weights based on
$\alpha$.

Results are presented in Table \ref{tab:mae-ci}. Averaged across
confidence levels and $n$ values, the test with the lowest $MAE$
remains the $AW(3,T)$ for the observed distribution, and the Iterative
Score for the simplex distribution. However, these averages are affected
by the higher $MAE$ seen in lower confidence levels. Results vary,
with tests which were generally liberal at 95\% performing better
at lower confidence levels, and vice versa. Averaging $MAE$ across
$n$, the best test for the 99\% level on both distributions is the
Iterative Score, followed by the $AW(3,T)$. For 80\% and 90\% confidence
levels, the best performing test is the Score $\left(\frac{2}{3}\right)$
for the observed distribution, and the $AW(z^{2},T)$ for the simplex.

\begin{table}

\includegraphics[width=6.5in]{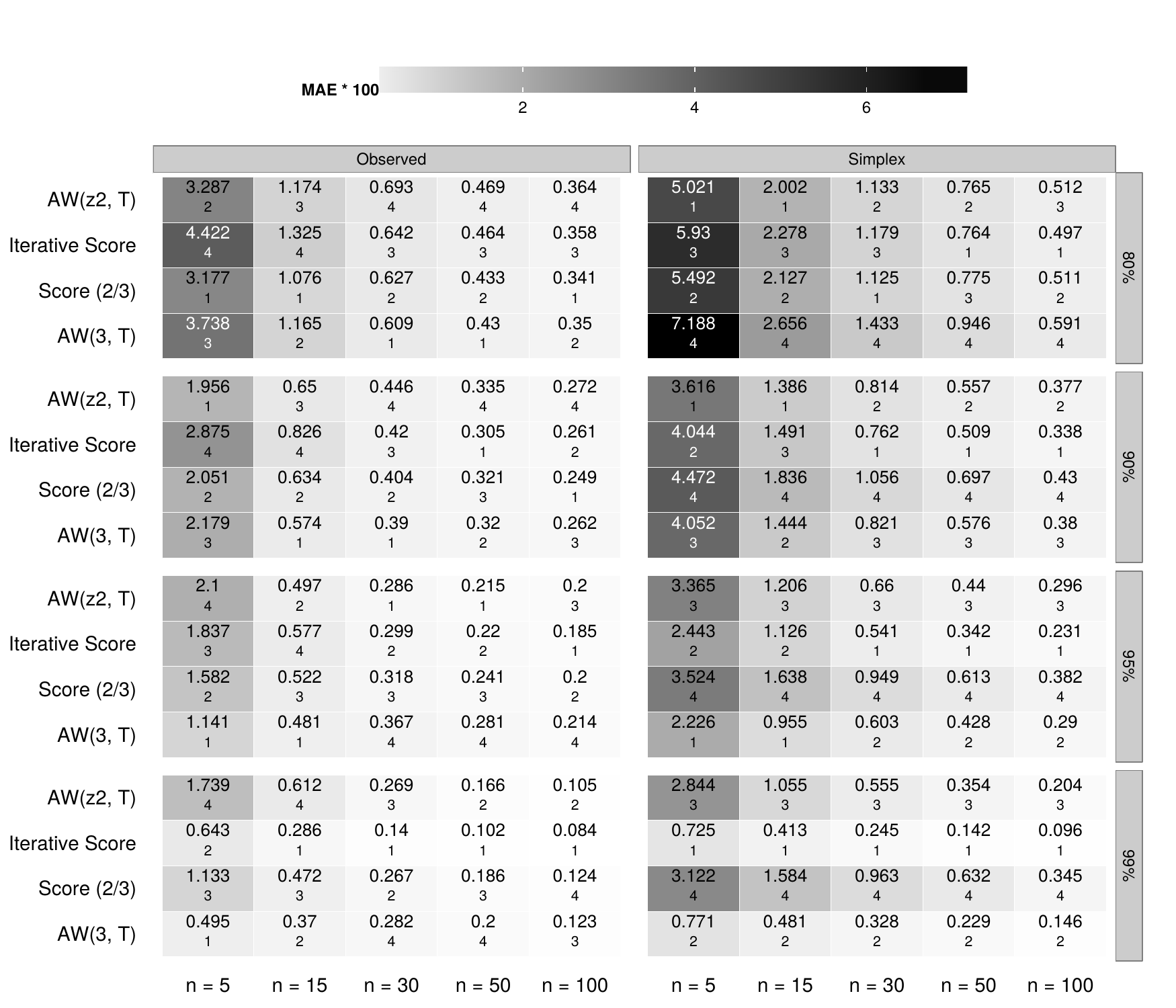}

\caption{Mean Absolute Error of estimation methods (multiplied by 100 for presentational
purposes) and rank, for the simplex and observed distributions across
confidence levels. Tests are ordered by descending total MAE for the
observed distribution, where $n\leqslant100$.\label{tab:mae-ci}}
\end{table}

\section{Conclusion \& Summary}

The Wald and Goodman tests perform poorly; their use should be avoided.
All Adjusted Wald variations considered provided substantial improvement,
with the best of those (weights of $3$ and $z_{\alpha/2}^{2}$) outperforming
non-iterative Score methods.

The best performing Adjusted Wald is $AW(3,T)$ which can be used
by adding $\frac{3}{4}$ to the counts of both \emph{Promoters} and
\emph{Detractors}, and $\frac{3}{2}$ to the count of \emph{Passives,}
before construction of a Wald interval. The method has good performance
across the $n$ values and confidence levels examined, especially
for data likely to be observed in practice.

The Iterative Score method also has excellent performance, with the
advantage that it has very few regions of parameter space where coverage
drops below 95\%, providing accurate coverage probabilities for almost
any trinomial distribution. Its disadvantage is its greater computational
complexity, and slight conservatism at low $n$ values ($<20$ for
a 95\% interval) for trinomial distributions likely to be observed
in practice.

\section{Acknowledgements}

The author would like to thank two anonymous reviewers and an assistant
editor for comments which greatly helped improve this manuscript.

\section{Trademark Information}

Net Promoter, Net Promoter Score, and NPS are trademarks of Satmetrix
Systems, Inc., Bain \& Company, Inc., and Fred Reichheld.

\bibliography{nps_interval_estimation.bbl}

\end{document}